\documentstyle[12pt]{article}
\renewcommand{\theequation}{\thesection.\arabic{equation}}

\newlength{\extraspace}
\newlength{\extraspaces}
\newcommand{\be}{\begin{equation}
\addtolength{\abovedisplayskip}{\extraspaces}
\addtolength{\belowdisplayskip}{\extraspaces}
\addtolength{\abovedisplayshortskip}{\extraspace}
\addtolength{\belowdisplayshortskip}{\extraspace}}
\newcommand{\ee}{\end{equation}}
\newcommand{\ba}{\begin{eqnarray}
\addtolength{\abovedisplayskip}{\extraspaces}
\addtolength{\belowdisplayskip}{\extraspaces}
\addtolength{\abovedisplayshortskip}{\extraspace}
\addtolength{\belowdisplayshortskip}{\extraspace}}
\newcommand{\ea}{\end{eqnarray}}
\newcommand{\newsection}[1]{
\vspace{7mm}
\pagebreak[3]
\addtocounter{section}{1}
\setcounter{equation}{0}
\setcounter{subsection}{0}

\begin{flushleft}
{\large {\bf \thesection. #1}}
\end{flushleft}
\nopagebreak
\medskip
\nopagebreak
\hspace{3mm}}

\renewcommand{\appendix}[3]{
\renewcommand{\thesection}{\Alph{section}}
\renewcommand{\theequation}{\thesection.\arabic{equation}}
\vspace{7mm}
\pagebreak[3]
\setcounter{section}{#2}
\setcounter{subsection}{0}
\setcounter{equation}{0}
\setcounter{footnote}{0}
\begin{flushleft}
{\large {\bf #1 #3}} 
\end{flushleft}
\nopagebreak
\medskip
\nopagebreak
\hspace{3mm}}
\newcommand{\nonu}{\nonumber \\[.5mm]}
\newcommand{\A}{&\!\!\!}

\begin{document}
\addtolength{\baselineskip}{.7mm}
\begin{flushright}
STUPP-96-149 \\ October, 1996
\end{flushright}
\vspace{.6cm}
\begin{center}
{\large{\bf{Structure of the field equations \\[2mm]
            in $N = 1$ chiral supergravity}}} \\[20mm]
{\sc Motomu Tsuda and Takeshi Shirafuji} \\[12mm]
Physics Department, Saitama University \\[2mm]
Urawa, Saitama 338, Japan \\[20mm]

{\parbox{13cm}{{\bf Abstract.} 
We study the structure of linearized field equations 
in $N = 1$ chiral supergravity (SUGRA) with a complex tetrad, 
as a preliminary to introducing additional auxiliary fields 
in order that the supersymmetry (SUSY) algebra close 
off shell. We follow the first-order formulation 
we have recently constructed using the method 
of the usual $N = 1$ SUGRA. 
In particular, we see how the real and imaginary parts 
of the complex tetrad are coupled to matter fields 
in the weak field approximation.
Starting from the linearized (free) theory 
of $N = 1$ chiral SUGRA, we then construct a Lagrangian 
which is invariant under local SUSY transformations 
to zeroth order of the gravitational constant, 
and compare the results with the linearized field equations.}} 
\end{center}
\vfill

\newpage
\setcounter{section}{0}
\setcounter{equation}{0}
\newsection{Introduction}

Right- and left-handed supersymmetry (SUSY) 
transformations in $N = 1$ chiral supergravity (SUGRA) 
with a complex tetrad were introduced by Jacobson 
\cite{AAN,JJ}, and their first-order formulation 
was then constructed using the two-form gravity 
\cite{CDJ,KS}. In a previous paper \cite{TS}, 
following the method used in the usual $N = 1$ SUGRA 
\cite{FN,DZ}, we presented the explicit form of 
the first-order SUSY transformations 
in $N = 1$ chiral SUGRA for complex field variables; 
a complex tetrad ${e^i}_{\mu}$, 
a self-dual connection $A^{(+)}_{ij \mu} 
= A^{(+)}_{[ij] \mu}$ which satisfies 
$(1/2){\epsilon_{ij}} \! ^{kl} A^{(+)}_{kl \mu} 
= i A^{(+)}_{ij \mu}$, 
and two independent (Majorana) Rarita-Schwinger 
fields $\psi_{R \mu}$ 
and $\overline{\tilde \psi}_{R \mu}$. 
\footnote{\ We assume $\psi_{\mu}$ 
and $\tilde \psi_{\mu}$ to be two independent 
(Majorana) Rarita-Schwinger fields, 
and define the right-handed spinor fields 
$\psi_{R \mu} := (1/2)(1 + \gamma_5) \psi_{\mu}$ 
and $\tilde \psi_{R \mu}:= (1/2)(1 + \gamma_5) 
\tilde \psi_{\mu}$. 
The $\psi_{R \mu}$ and $\tilde \psi_{R \mu}$ 
relate to the left-handed spinor fields 
$\psi_{L \mu}$ and $\tilde \psi_{L \mu}$ respectively, 
because $\psi_{\mu}$ and $\tilde \psi_{\mu}$ 
are Majorana spinors. 
The antisymmetrization of a tensor with respect to $i$ 
and $j$ is denoted by $A_{[i \mid \cdots \mid j]} 
:= (1/2)(A_{i \cdots j} - A_{j \cdots i})$. 
We shall follow the notation and convention 
of \cite{TS}.}
The SUSY transformation parameters are not 
constrained at all in contrast with the method of 
the two-form gravity. 
We showed that the SUSY algebra for $N = 1$ chiral SUGRA 
closes only on shell. 

The sum of the right- and left-handed SUSY transformations, 
however, is not twice the usual $N = 1$ SUGRA 
in the following sense: 
The sum of the right- and left-handed SUSY transformations 
for the complex tetrad is 
\be
\delta {e^i}_{\mu} 
      = -i(\overline{\tilde \psi}_{R \mu} \gamma^i \alpha_R 
           + \overline \psi_{L \mu} \gamma^i \tilde \alpha_L) 
\label{SUSY-e}
\ee
with $\alpha$ and $\tilde \alpha$ being two anticommuting 
Majorana spinor parameters. 
Here both $\overline{\tilde \psi}_{R \mu} \gamma^i \alpha_R$ 
and $\overline \psi_{L \mu} \gamma^i \tilde \alpha_L$ 
are complex. 
The real and imaginary parts of (\ref{SUSY-e}) 
cannot be written as the form 
\be
{\rm Re}(\delta {e^i}_{\mu}) 
= -i \overline \Phi^1_{\mu} \gamma^i \beta^1, \ \ 
{\rm Im}(\delta {e^i}_{\mu}) 
= -i \overline \Phi^2_{\mu} \gamma^i \beta^2 
\label{SUSY-e!}
\ee
by field redifinition from $\psi_{\mu}$ and $\tilde \psi_{\mu}$ 
to appropriate Majorana spinors $\Phi^1_{\mu}$ and $\Phi^2_{\mu}$, 
accompanied by corresponding change of 
$\alpha$ and $\tilde \alpha$ to Majorana spinor 
parameters $\beta^1$ and $\beta^2$. 
Instead, we have 
\ba
\A \A {\rm Re}(\delta {e^i}_{\mu}) 
      = -{i \over 2} (\overline{\tilde \psi}_{\mu} \gamma^i \alpha 
           + \overline \psi_{\mu} \gamma^i \tilde \alpha), 
\label{SUSY-re} \\
\A \A {\rm Im}(\delta {e^i}_{\mu}) 
      = -{i \over 2} (\overline{\tilde \psi}'_{\mu} \gamma^i \alpha 
           + \overline \psi'_{\mu} \gamma^i \tilde \alpha), 
\label{SUSY-ie}
\ea
where $\overline{\tilde \psi}'_{\mu} := 
\overline{\tilde \psi}_{\mu} 
{\rm exp} \{ (-i \pi \gamma_5)/2 \}$ 
and $\overline \psi'_{\mu} := \overline \psi_{\mu} 
{\rm exp} \{ (i \pi \gamma_5)/2 \}$. 
Both (\ref{SUSY-re}) and (\ref{SUSY-ie}) contain 
two kinds of Majorana spinor parameters. 

Therefore it seems non-trivial to introduce additional 
auxiliary fields which will make the SUSY algebra 
of $N = 1$ chiral SUGRA closed off shell. 
The full non-linear theory with auxiliary fields 
of the usual $N = 1$ SUGRA can be constructed 
from its linearized theory, 
making the rigid SUSY transformations local 
and adding appropriate terms to the free Lagrangian 
order-by-order in the gravitational constant $\kappa$ 
\cite{West}. 
\footnote{\ The $\kappa^2$ is the Einstein constant: 
$\kappa^2 = 8 \pi G/c^4$. Unless stated otherwise, 
we use units $c = 1 = \kappa^2$.}
This suggests that if we can introduce additional auxiliary fields 
at linearized level, the full nonlinear theory 
with auxiliary fields will be constructed 
also for $N = 1$ chiral SUGRA. 
Motivated by this expectation, we consider the structure 
of the linearized field equations in $N = 1$ chiral SUGRA. 
In particular, we see how the real and imaginary parts 
of the complex tetrad are coupled to matter fields 
in the weak field approximation. 
We shall then modify the linearized (free) Lagrangian 
so that it be invariant under local SUSY transformations 
up to order $\kappa^0$, 
and show that the modified Lagrangian correctly 
reproduces the field equations of $N = 1$ chiral SUGRA 
in the weak field approximation. 

This paper is organized as follows. In section 2 
we define a real Lagrangian from chiral one 
which is assumed to be analytic in the complex field variables, 
and derive the field equations for the real 
and imaginary parts of the complex tetrad. 
The chiral Lagrangian of matter fields includes 
massless Majorana spin-1/2 and spin-3/2 fields. 
In section 3 we apply the weak field approximation 
to the field equations derived from the real Lagrangian. 
The explicit form of energy-momentum tensors 
is calculated for (Majorana) Rarita-Schwinger fields. 
In section 4 we construct a local SUSY invariant Lagrangian 
to order $\kappa^0$, and compare the resultant field equations 
with the linearized field equations in $N = 1$ chiral SUGRA. 
In section 5 we present our conclusion. 
We summarize the identities derived from general coordinate 
and local Lorentz invariances of the chiral Lagrangian 
in the appendix.

\newpage
\newsection{Real Lagrangian}

In local field theory, the spinor field $\psi$ usually 
appears in the kinetic Lagrangian forming a bilinear product 
with its own Dirac conjugate $\overline \psi$. 
However, in $N = 1$ chiral SUGRA, 
we must use the kinetic Lagrangian formed 
of the bilinear product of the two independent spinor 
fields $\psi_{R \mu}$ and $\overline{\tilde \psi}_{R \mu}$ 
in order to make the SUSY transformations 
compatible with the complex tetrad. 
Therefore, to recover the Hermiticity of the kinetic Lagrangian, 
we add the complex conjugate, $\overline{{\cal L}^{(+)}}$, 
to the chiral Lagrangian density ${\cal L}^{(+)}$: 
Namely, we define the real Lagrangian density, 
\footnote{\ For pure gravity case, 
Peld\'{a}n \cite{PP} used the real Lagrangian to derive 
the real Ashtekar-like theory based on the Lie-algebra 
$SO(3,1)$, denoting the real Lagrangian density 
by ${\cal L}^{{\rm tot}}$.}
\be
{\cal L} := {\cal L}^{(+)} + {\rm c.c.}, 
\label{Lreal}
\ee
where ${\cal L}^{(+)}$ is the sum of the chiral 
gravitational Lagrangian density and the chiral 
Lagrangian density of matter fields, 
and ``c.c.'' means ``the complex conjugate 
of the preceding term''. 
The chiral gravitational Lagrangian density 
constructed from the complex tetrad 
and the self-dual connection is 
\be
{\cal L}^{(+)}_G = -{i \over 2} e \ \epsilon^{\mu \nu \rho \sigma} 
   {e^i}_{\mu} {e^j}_{\nu} R^{(+)}_{ij \rho \sigma}, 
\label{LG+}
\ee
where $e$ denotes ${\rm det}(e^i_{\mu})$ 
and the curvature of self-dual connection 
${R^{(+)ij}}_{\mu \nu}$ is 
\be
{R^{(+)ij}}_{\mu \nu} := 2(\partial_{[\mu} {A^{(+)ij}}_{\nu]} 
             + {A^{(+)i}}_{k [\mu} {A^{(+)kj}}_{\nu]}). 
\label{curv+}
\ee

In order to discuss the structure of linearized field equations 
as generally as possible, we suppose that 
the chiral Lagrangian density of matter fields take the form 
\be
{\cal L}^{(+)}_M := {\cal L}^{(+)}_M [e, A^{(+)}, 
\overline{\tilde \Psi}_R, D^{(+)}_\mu \Psi_R], 
\label{LM+}
\ee
which is analytic in the complex field variables, 
${e^i}_{\mu}$, $A^{(+)}_{ij \mu}$, $\Psi_R$ 
and $\overline{\tilde \Psi}_R$. 
Here $\Psi_R$ and $\overline{\tilde \Psi}_R$ 
are independent of each other, and denote collectively 
the matter fields; the $D^{(+)}_\mu$ means 
the covariant derivative with respect to $A^{(+)}_{ij \mu}$: 
\be
D^{(+)}_\mu := \partial_\mu 
              + {i \over 2} A^{(+)}_{ij \mu} S^{ij}. 
\ee
The ${\cal L}^{(+)}_M$ of (\ref{LM+}) is invariant 
under general coordinate and local Lorentz transformations, 
but its SUSY invariance is not necessarily satisfied. 

The matter fields under consideration include massless 
Majorana spin-1/2 and spin-3/2 fields. 
For massless Majorana spin-1/2 fields, 
the chiral Lagrangian density is 
\be
{\cal L}^{(+)}_{1/2} = {e \over 6} \ \epsilon^{\mu \nu \rho \sigma} 
                   \overline{\tilde \psi}_R \gamma_{\nu \rho \sigma} 
                   D^{(+)}_\mu \psi_R 
\ee
with $\gamma_{\nu \rho \sigma} 
:= \gamma_{[\nu} \gamma_{\rho} \gamma_{\sigma]}$. 
\footnote{The totally antisymmetrization 
of a tensor with respect to $i, j$ and $k$ is
denoted by $A_{[ijk]} := (1/3)(A_{i[jk]} 
+ A_{k[ij]} + A_{j[ki]})$.}
For (Majorana) Rarita-Schwinger fields, we have 
\be
{\cal L}^{(+)}_{{\rm RS}} = - e \ \epsilon^{\mu \nu \rho \sigma} 
                     \overline{\tilde \psi}_{R \mu} \gamma_\rho 
                     D^{(+)}_\sigma \psi_{R \nu}. 
\ee

Note that the ${\cal L}^{(+)}_M$ of (\ref{LM+}) 
changes by phase under global chiral transformations, 
\ba
\Psi_R(x) \A \rightarrow \A {\rm exp}(i \gamma_5 \theta) \Psi_R(x) 
                            = {\rm exp}(i \theta) \Psi_R(x), 
\label{P-ctr} \\
\overline{\tilde \Psi}_R(x) \A \rightarrow \A 
         \overline{\tilde \Psi}_R(x) {\rm exp}(i \gamma_5 \theta') 
         = \overline{\tilde \Psi}_R(x) {\rm exp}(-i \theta'), 
\label{Pt-ctr}
\ea
where $\theta$ and $\theta'$ are real constant parameters. 
Accordingly, we define the chiral Lagrangian density 
${\cal L}^{(+)}$ as 
\be
{\cal L}^{(+)} = {\cal L}^{(+)}_G 
+ e^{i \varphi} {\cal L}^{(+)}_{M}, 
\ee
with $\varphi$ being a real constant parameter. 
Then $\varphi$ is changed like $\varphi \rightarrow 
\varphi + (\theta - \theta')$ under the chiral transformations 
(\ref{P-ctr}) and (\ref{Pt-ctr}). 

Let us express the real gravitational Lagrangian density, 
\be
{\cal L}_G := {\cal L}^{(+)}_G + {\rm c.c.}, 
\label{LGreal-0}
\ee
by using only real variables. 
We define the real $SO(3,1)$ connection by 
\be
\omega_{ij \mu} := A^{(+)}_{ij \mu} + {\rm c.c.} 
\ee
Then the ${R^{(+)ij}}_{\mu \nu}$ of (\ref{curv+}) becomes 
\be
{R^{(+)ij}}_{\mu \nu} 
= {1 \over 2} \left({R^{ij}}_{\mu \nu}[\omega] 
  - {i \over 2} {\epsilon^{ij}} \! _{kl} 
  {R^{kl}}_{\mu \nu}[\omega] \right), 
\ee
where ${R^{ij}}_{\mu \nu}[\omega]$ 
is the curvature of the real connection $\omega_{ij \mu}$. 
Further we decompose the complex tetrad 
into the real and imaginary parts: 
\be
{e^i}_{\mu} = {V^i}_{\mu} + i {W^i}_{\mu}. 
\ee
The ${\cal L}_G$ of (\ref{LGreal-0}) can then be written as 
\be
{\cal L}_G 
= -{e \over 4} \ \epsilon^{\mu \nu \rho \sigma} 
  ({V^i}_{\mu} {V^j}_{\nu} - {W^i}_{\mu} {W^j}_{\nu}) 
  {\epsilon_{ij}} \! ^{kl} R_{kl \rho \sigma}[\omega] 
  + e \ \epsilon^{\mu \nu \rho \sigma} 
   {V^i}_{\mu} {W^j}_{\nu} R_{ij \rho \sigma}[\omega]. 
\label{LGreal}
\ee

The real Lagrangian density of matter fields, 
${\cal L}_M$, can be written by using 
${V^i}_{\mu}$, ${W^i}_{\mu}$ and $\omega_{ij \mu}$ as 
\ba
{\cal L}_M 
\A := \A e^{i \varphi} {\cal L}^{(+)}_M + {\rm c.c.} \nonu
\A = \A {\cal L}_M [V, W, \omega, 
\overline{\tilde \Psi}, D_{\mu}[\omega] \Psi], 
\label{LMreal}
\ea
where $D_{\mu}[\omega]$ denotes 
the covariant derivative 
with respect to $\omega_{ij \mu}$: 
\be
D_{\mu}[\omega] := \partial_\mu 
                  + {i \over 2} \omega_{ij \mu} S^{ij}. 
\ee
Note that $\overline{\tilde \Psi}$ and $\Psi$ appear 
in (\ref{LMreal}) instead of 
$\overline{\tilde \Psi}_R$ and $\Psi_R$. 
In fact, the real Lagrangian density of 
massless Majorana spin-1/2 fields 
can be expressed as 
\ba
{\cal L}_{1/2} \A = \A 
{e \over 6} \ \epsilon^{\mu \nu \rho \sigma} 
\{ {V^{ijk}} \! _{\nu \rho \sigma} 
(e^{i \varphi} \overline{\tilde \psi}_R \gamma_{ijk} 
D_{\mu}[\omega] \psi_R 
- e^{-i \varphi} \overline{\tilde \psi}_L \gamma_{ijk} 
D_{\mu}[\omega] \psi_L) \nonu
\A \A + {W^{ijk}} \! _{\nu \rho \sigma} 
      (e^{i (\varphi + {\pi \over 2})} 
      \overline{\tilde \psi}_R \gamma_{ijk} 
      D_{\mu}[\omega] \psi_R 
      - e^{-i (\varphi + {\pi \over 2})} 
      \overline{\tilde \psi}_L \gamma_{ijk} 
      D_{\mu}[\omega] \psi_L) \} 
\label{L1/2real}
\ea
with ${V^{ijk}} \! _{\nu \rho \sigma}$ 
and ${W^{ijk}} \! _{\nu \rho \sigma}$ being defined by 
\ba
{V^{ijk}} \! _{\nu \rho \sigma} 
\A := \A {V^{[i}}_{[\nu} ({V^j}_{\rho} {V^{k]}}_{\sigma]} 
         - 3 {W^j}_{\rho} {W^{k]}}_{\sigma]}), \\
{W^{ijk}} \! _{\nu \rho \sigma} 
\A := \A {W^{[i}}_{[\nu} (3 {V^j}_{\rho} {V^{k]}}_{\sigma]} 
         - {W^j}_{\rho} {W^{k]}}_{\sigma]}). 
\ea
For massless (Majorana) Rarita-Schwinger fields, 
the real Lagrangian density is 
\ba
{\cal L}_{{\rm RS}} \A = \A 
- e \ \epsilon^{\mu \nu \rho \sigma} \{ {V^i}_{\rho} 
(e^{i \varphi} \overline{\tilde \psi}_{R \mu} \gamma_i 
D_{\sigma}[\omega] \psi_{R \nu} 
- e^{-i \varphi} \overline{\tilde \psi}_{L \mu} \gamma_i 
D_{\sigma}[\omega] \psi_{L \nu}) \nonu
\A \A + {W^i}_{\rho} 
      (e^{i (\varphi + {\pi \over 2})} 
      \overline{\tilde \psi}_{R \mu} \gamma_i 
      D_{\sigma}[\omega] \psi_{R \nu} 
      - e^{-i (\varphi + {\pi \over 2})} 
      \overline{\tilde \psi}_{L \mu} \gamma_i 
      D_{\sigma}[\omega] \psi_{L \nu}) \}. 
\label{LRSreal}
\ea

>From (\ref{LGreal}) and (\ref{LMreal}), 
we can derive the field equations for ${V^i}_{\mu}$, 
${W^i}_{\mu}$ and $\omega_{ij \mu}$. 
Varying ${\cal L} = {\cal L}_G + {\cal L}_M$ 
with respect to ${V^i}_{\mu}$ and ${W^i}_{\mu}$ 
yields 
\ba
\A \A e \ \epsilon^{\mu \nu \rho \sigma} 
      ({W^j}_{\nu} R_{ij \rho \sigma}[\omega] - {1 \over 2} 
      {V^j}_{\nu} {\epsilon_{ij}} \! ^{kl} R_{kl \rho \sigma}[\omega]) 
      + e \ {{T^{(1)}}_i}^{\mu} = 0, 
\label{eq-V} \\
\A \A e \ \epsilon^{\mu \nu \rho \sigma} 
      ({V^j}_{\nu} R_{ij \rho \sigma}[\omega] + {1 \over 2} 
      {W^j}_{\nu} {\epsilon_{ij}} \! ^{kl} R_{kl \rho \sigma}[\omega]) 
      + e \ {{T^{(2)}}_i}^{\mu} = 0, 
\label{eq-W}
\ea
respectively. 
We note that there appear two energy-momentum tesors 
of matter fields, ${{T^{(1)}}_i}^{\mu}$ and ${{T^{(2)}}_i}^{\mu}$, 
defined by 
\be
{{T^{(1)}}_i}^{\mu} := e^{-1} \ 
{{\delta {\cal L}_M} \over {\delta {V^i}_{\mu}}}, \ \ \ 
{{T^{(2)}}_i}^{\mu} := e^{-1} \ 
{{\delta {\cal L}_M} \over {\delta {W^i}_{\mu}}}. 
\ee
These are related to 
the complex energy-momentum tensor, 
\be
{{T^{(+)}}_i}^{\mu}:= e^{-1} \ 
{{\delta {\cal L}^{(+)}_M} \over {\delta {e^i}_{\mu}}}, 
\label{EM+}
\ee
by 
\be
{{T^{(1)}}_i}^{\mu} - i{{T^{(2)}}_i}^{\mu} 
= 2 {{T^{(+)}}_i}^{\mu} 
\label{rel-T}
\ee
due to the Cauchy-Riemann relation. 
For $\omega_{ij \mu}$, we have the field equation, 
\be
e \ \epsilon^{\mu \nu \rho \sigma} 
D_{\rho}[\omega] {\rm Im}{H^{(+)ij}}_{\mu \nu} 
+ {e \over 2} \ S^{ij \sigma} = 0, 
\label{eq-o}
\ee
where the spin tensor of matter fields is defined by 
\be
S^{ij \mu} := -2 e^{-1} \ 
{{\delta {\cal L}_M} \over {\delta \omega_{ij \mu}}}, 
\label{spin}
\ee
and ${\rm Im}{H^{(+)ij}}_{\mu \nu}$ is the imaginary 
part of ${H^{(+)ij}}_{\mu \nu} := {e^i}_{[\mu} {e^j}_{\nu]} 
- (i/2) {\epsilon^{ij}} \! _{kl} 
{e^k}_{[\mu} {e^l}_{\nu]}$. 
Note also that the $S^{ij \mu}$ of (\ref{spin}) 
is connected with the self-dual spin tensor, 
\be
S^{(+)ij \mu}:= -2 e^{-1} \ 
{{\delta {\cal L}^{(+)}_M} \over {\delta A^{(+)}_{ij \mu}}}, 
\label{spin+}
\ee
by 
\be
S^{ij \mu} = S^{(+)ij \mu} + {\rm c.c.} 
\label{rel-S}
\ee
The relation between ${{T^{(+)}}_i}^{\mu}$ 
and $S^{(+)ij \mu}$ is shown in the appendix 
(see (\ref{Nid-L1})). 

If we impose the reality condition, 
${W^i}_{\mu} = 0$ and 
$\overline{\tilde \Psi} = \overline \Psi$, 
then the two real constant parameters $\theta$ and $\theta'$ 
in (\ref{P-ctr}) and (\ref{Pt-ctr}) 
are equal to each other, i.e., $\theta = \theta'$, 
so the ${\cal L}^{(+)}_M$ of (\ref{LM+}) 
becomes strictly invariant under the global chiral transformations. 
Thus we choose the phase factor $\varphi = 0$. 
Then, in the case of spin-1/2 and spin-3/2 fields, 
substituting the solution of (\ref{eq-o}) 
into (\ref{eq-V}) and (\ref{eq-W}) 
yields the ordinary Einstein equation 
and the Bianchi identity respectively.

\newsection{Weak field approximation}

To see how the real and imaginary parts 
of the complex tetrad are coupled to matter fields, 
we apply the weak field approximation 
to the field equations (\ref{eq-V}), (\ref{eq-W}) 
and (\ref{eq-o}), 
assuming that ${V^i}_{\mu}$ and ${W^i}_{\mu}$ satisfy 
\ba
\A \A {V^i}_{\mu} = {\delta^i}_{\mu} + {a^i}_{\mu}, \ \ 
      |{a^i}_{\mu}| \ll 1, \\
\A \A {W^i}_{\mu} = {b^i}_{\mu}, \ \ 
      |{b^i}_{\mu}| \ll 1. 
\ea
It is convenient to decompose the self-dual connection 
$A^{(+)}_{ij \mu}$ as 
\be
A^{(+)}_{ij \mu} = A^{(+)}_{ij \mu}[e] 
+ K^{(+)}_{ij \mu} 
\ee
and to take $K^{(+)}_{ij \mu}$ as an independent variable 
instead of $A^{(+)}_{ij \mu}$. 
Here $A^{(+)}_{ij \mu}[e]$ is the self-dual 
part of the Ricci rotation coefficients $A_{ij \mu}[e]$. 
Since we need not distinguish Latin indices from Greek ones 
in the weak field approximation, 
we use Greek indices in the rest of this paper, 
which are raised and lowered with the Minkowski 
metric tensor $\eta_{\mu \nu}$. 
Now the real $SO(3,1)$ connection $\omega_{\mu \nu \lambda}$ 
can be expressed in terms of $a_{\mu \nu}$, $b_{\mu \nu}$ 
and $K^{(+)}_{\mu \nu \lambda}$ as follows: 
\be
\omega_{\mu \nu \lambda} = \omega_{\mu \nu \lambda}[a] 
+ {1 \over 2} {\epsilon_{\mu \nu}} \! ^{\rho \sigma} 
\omega_{\rho \sigma \lambda}[b] 
+ K_{\mu \nu \lambda}, 
\ee
where 
\ba
\omega_{\mu \nu \lambda}[a] 
     \A := \A - \partial_{\lambda} a_{[\mu \nu]} 
     - (\partial_{\mu} a_{(\nu \lambda)} 
     - \partial_{\nu} a_{(\mu \lambda)}), \\
\omega_{\mu \nu \lambda}[b] 
     \A := \A - \partial_{\lambda} b_{[\mu \nu]} 
     - (\partial_{\mu} b_{(\nu \lambda)} 
     - \partial_{\nu} b_{(\mu \lambda)}), 
\ea
and $K_{\mu \nu \lambda}$ is defined by 
\be
K_{\mu \nu \lambda} 
:= K^{(+)}_{\mu \nu \lambda} + {\rm c.c.} 
\ee
Therefore, the curvature $R_{\mu \nu \rho \sigma}[\omega]$ 
can be written as 
\be
R_{\mu \nu \rho \sigma}[\omega] 
= R_{\mu \nu \rho \sigma}[a] 
  + {1 \over 2} {\epsilon_{\mu \nu}} \! ^{\alpha \beta} 
  R_{\alpha \beta \rho \sigma}[b] 
  + 2 \partial_{[\rho} K_{|\mu \nu| \sigma]}, 
\label{curv-o}
\ee
where 
\ba
R_{\mu \nu \rho \sigma}[a] 
\A := \A 2 \partial_{[\rho} 
         \omega_{|\mu \nu| \sigma]}[a], \\
R_{\mu \nu \rho \sigma}[b] 
\A := \A 2 \partial_{[\rho} 
         \omega_{|\mu \nu| \sigma]}[b], 
\ea
which are linear in $a_{(\mu \nu)}$ and $b_{(\mu \nu)}$ 
respectively. 

Substituting (\ref{curv-o}) into (\ref{eq-V}) and (\ref{eq-W}), 
we can get the field equations in the weak field approximation. 
In (\ref{eq-V}), the term of $e \ \epsilon^{\mu \nu \rho \sigma} 
{W^j}_{\nu} R_{ij \rho \sigma}[\omega]$ 
can be neglected, and the term proportional to 
$R_{\mu \nu \rho \sigma}[b]$ vanishes 
due to the Bianchi identity. Therefore (\ref{eq-V}) becomes 
\be
G_{\mu \nu}[a] + \partial^{\rho} K_{\rho \nu \mu} 
+ \partial_{\mu} v_{\nu} 
- \eta_{\mu \nu} \partial^{\rho} v_{\rho} 
= {1 \over 2} T^{(1)}_{\mu \nu} 
\label{leq-a}
\ee
with $v_{\mu} := {K_{\mu \nu}}^{\nu}$. 
Here $G_{\mu \nu}[a]$ is the linearized Einstein tensor 
for $a_{\mu \nu}$, 
\be
G_{\mu \nu}[a] = - \{ \Box {\overline a_{(\mu \nu)}} 
   - \partial^{\rho} (\partial_{\mu} {\overline a_{(\nu \rho)}} 
   + \partial_{\nu} {\overline a_{(\mu \rho)}}) 
   + \eta_{\mu \nu} \partial^{\rho} \partial^{\sigma} 
   {\overline a_{(\rho \sigma)}} \} 
\ee
with 
\be
{\overline a_{(\mu \nu)}} := a_{(\mu \nu)} 
- {1 \over 2} \eta_{\mu \nu} a, 
\ \ a := \eta^{\mu \nu} a_{(\mu \nu)}, 
\ee
and the d'Alembertian $\Box$ being defined 
by $\Box := \partial^{\mu} \partial_{\mu}$. 
Similarly, from (\ref{eq-W}) we have 
\be
G_{\mu \nu}[b] 
+ \epsilon_{\lambda \rho \sigma \nu} 
\partial^{\lambda} {K_{\mu}}^{\rho \sigma} 
= - {1 \over 2} T^{(2)}_{\mu \nu}, 
\label{leq-b}
\ee
where $G_{\mu \nu}[b]$ is the linearized Einstein tensor 
for $b_{\mu \nu}$. 
Further, the field equation (\ref{eq-o}) becomes 
\be
- K_{\rho [\mu \nu]} + \eta_{\rho [\mu} v_{\nu]} 
= {1 \over 4} S_{\mu \nu \rho}. 
\label{leq-o}
\ee

We take the energy-momentum tensors, $T^{(1)}_{\mu \nu}$ and 
$T^{(2)}_{\mu \nu}$, 
and the spin tensor $S_{\mu \nu \rho}$ 
to lowest order in  $a_{\mu \nu}$, $b_{\mu \nu}$ 
and $K_{\mu \nu \rho}$: 
Namely, they are independent of $a_{\mu \nu}$, $b_{\mu \nu}$ 
and $K_{\mu \nu \rho}$, 
and satisfy the conservation law, 
\be
\partial^{\nu} T^{(1)}_{\mu \nu} = 0, \ \ 
\partial^{\nu} T^{(2)}_{\mu \nu} = 0, 
\label{conserv}
\ee
and the Tetrode formula in special relativity, 
\be
T^{(1)}_{[\mu \nu]} 
- {1 \over 2} {\epsilon_{\mu \nu}} \! ^{\rho \sigma} 
T^{(2)}_{\rho \sigma} 
= \partial^{\rho} S_{\mu \nu \rho}, 
\label{Tet}
\ee
as is shown in the appendix 
(see (\ref{Tet+}) and (\ref{conserv+})). 
Note that the antisymmetric parts of 
$T^{(1)}_{\mu \nu}$ and $T^{(2)}_{\mu \nu}$ 
are related with each other by 
\be
2 T^{(1)}_{[\mu \nu]} 
= - {\epsilon_{\mu \nu}} \! ^{\rho \sigma} 
T^{(2)}_{\rho \sigma}. 
\label{T1-T2}
\ee

The linearized field equations (\ref{leq-a}), 
(\ref{leq-b}) and (\ref{leq-o}) 
are invariant under the gauge transformations, 
\ba
\A \A a_{(\mu \nu)} \rightarrow a'_{(\mu \nu)} 
           = a_{(\mu \nu)} + \partial_{\mu} \Lambda^1_{\nu} 
           + \partial_{\nu} \Lambda^1_{\mu}, 
\label{lgtr-a} \\
\A \A a_{[\mu \nu]} \rightarrow a'_{[\mu \nu]} 
           = a_{[\mu \nu]} + \epsilon^1_{\mu \nu}, 
\label{lLtr-a}
\ea
and 
\ba
\A \A b_{(\mu \nu)} \rightarrow b'_{(\mu \nu)} 
           = b_{(\mu \nu)} + \partial_{\mu} \Lambda^2_{\nu} 
           + \partial_{\nu} \Lambda^2_{\mu}, 
\label{lgtr-b} \\
\A \A b_{[\mu \nu]} \rightarrow b'_{[\mu \nu]} 
           = b_{[\mu \nu]} + \epsilon^2_{\mu \nu} 
\label{lLtr-b}
\ea
where $\Lambda^a_{\mu}$ and $\epsilon^a_{\mu \nu} = 
\epsilon^a_{[\mu \nu]} (a = 1, 2)$ 
are arbitrary four and six functions, respectively. 
These transformations are the linearized version 
of complex general coordinate 
and complex local Lorentz transformations. 
Equations (\ref{lLtr-a}) and (\ref{lLtr-b}) mean that 
$a_{[\mu \nu]}$ and $b_{[\mu \nu]}$ can be eliminated. 
By means of the gauge freedom we can put the harmonic condition, 
\be
\partial^{\nu} {\overline a_{(\mu \nu)}} = 0, \ \ 
\partial^{\nu} {\overline b_{(\mu \nu)}} = 0. 
\label{harm}
\ee
Then the remaining degrees of freedom are 6 
for each $a_{\mu \nu}$ and $b_{\mu \nu}$, 
and the linearized Einstein tensors for $a_{\mu \nu}$ 
and $b_{\mu \nu}$ are written as 
\be
G_{\mu \nu}[a] = - \Box {\overline a_{(\mu \nu)}}, \ \ 
G_{\mu \nu}[b] = - \Box {\overline b_{(\mu \nu)}}. 
\ee

Equation (\ref{leq-a}) is decomposed 
into the symmetric and antisymmetric parts as 
\ba
\A \A G_{\mu \nu}[a] + \partial^{\rho} K_{\rho (\mu \nu)} 
      + \partial_{(\mu} v_{\nu)} 
      - \eta_{\mu \nu} \partial^{\rho} v_{\rho} 
      = {1 \over 2} T^{(1)}_{(\mu \nu)}, 
\label{leq-asym} \\
\A \A - \partial^{\rho} K_{\rho [\mu \nu]} 
      + \partial_{[\mu} v_{\nu]} 
      = {1 \over 2} T^{(1)}_{[\mu \nu]}, 
\label{leq-aasym}
\ea
and similarly (\ref{leq-b}) into 
\ba
\A \A G_{\mu \nu}[b] 
      + \epsilon_{\lambda \rho \sigma (\mu} 
      \partial^{\lambda} {K_{\nu)}}^{\rho \sigma} 
      = - {1 \over 2} T^{(2)}_{(\mu \nu)}, 
\label{leq-bsym} \\
\A \A - \epsilon_{\lambda \rho \sigma [\mu} 
      \partial^{\lambda} {K_{\nu]}}^{\rho \sigma} 
      = - {1 \over 2} T^{(2)}_{[\mu \nu]}. 
\label{leq-basym}
\ea
Both sides of (\ref{leq-a}) and (\ref{leq-b}) 
are divergenceless with respect to $\nu$, 
and the divergence of (\ref{leq-o}) with respect to $\rho$ 
yields (\ref{leq-aasym}) and (\ref{leq-basym}) 
due to the Tetrode formula (\ref{Tet}). 
Therefore, there are $(16 + 16 + 24) - (4 + 4 + 6 + 6) = 36$ 
independent equations for $6 + 6 + 24 = 36$ 
independent variables, $a_{(\mu \nu)}$, $b_{(\mu \nu)}$ 
and $K_{\mu \nu \rho}$. 

We can rewrite (\ref{leq-asym}) and (\ref{leq-bsym}) 
with the help of (\ref{leq-o}). Taking the divergence 
of (\ref{leq-o}) with respect to $\mu$ yields 
\be
\partial^{\rho} K_{\rho (\mu \nu)} 
+ \partial_{(\mu} v_{\nu)} 
- \eta_{\mu \nu} \partial^{\rho} v_{\rho} 
= {1 \over 2} \partial^{\rho} S_{\rho (\mu \nu)}, 
\label{div-o1}
\ee
or equivalently 
\be
\epsilon_{\lambda \rho \sigma (\mu} 
\partial^{\lambda} {K_{\nu)}}^{\rho \sigma} 
= - {1 \over 2} \partial^{\rho} S^*_{\rho (\mu \nu)} 
\label{div-o2}
\ee
with the $\ast$ operation being the duality operation 
\be
S^*_{\mu \nu \rho} := {1 \over 2} 
{\epsilon_{\mu \nu}} \! ^{\alpha \beta} S_{\alpha \beta \rho}. 
\label{S-star}
\ee
Using (\ref{div-o1}) and (\ref{div-o2}) 
in (\ref{leq-asym}) and (\ref{leq-bsym}) respectively, 
we get 
\ba
\A \A G_{\mu \nu}[a] 
= {1 \over 2} T^{(1) ({\rm sym})}_{\mu \nu}, 
\label{leq-Ta} \\
\A \A G_{\mu \nu}[b] 
= - {1 \over 2} T^{(2) ({\rm sym})}_{\mu \nu}, 
\label{leq-Tb}
\ea
where $T^{(1) ({\rm sym})}_{\mu \nu}$ 
and $T^{(2) ({\rm sym})}_{\mu \nu}$ 
are the symmetrized energy-momentum tensors: 
\ba
T^{(1) ({\rm sym})}_{\mu \nu} \A = \A T^{(1)}_{\mu \nu} 
  - {1 \over 2} \partial^{\rho} 
  (S_{\mu \nu \rho} + S_{\rho \mu \nu} 
  + S_{\rho \nu \mu}), \\
T^{(2) ({\rm sym})}_{\mu \nu} \A = \A T^{(2)}_{\mu \nu} 
  - {1 \over 2} \partial^{\rho} 
  (S^*_{\mu \nu \rho} + S^*_{\rho \mu \nu} 
  + S^*_{\rho \nu \mu}). 
\label{T2-sym}
\ea

Now we restrict ourselves to the case 
of (Majorana) Rarita-Schwinger fields. 
>From the real Lagrangian density (\ref{LRSreal}), 
$T^{(1)}_{\mu \nu}$ and $T^{(2)}_{\mu \nu}$ can be obtained as 
\ba
\left( \matrix{ T^{(1)}_{\mu \nu} \cr
                T^{(2)}_{\mu \nu} \cr} \right) 
= \left( \matrix{ {\rm cos}\varphi & {\rm sin}\varphi \vspace{2mm} \cr
                 -{\rm sin}\varphi & {\rm cos}\varphi \cr} \right) 
  \left( \matrix{ T^{(1)}_{\mu \nu}|_{\varphi = 0} \cr
                  T^{(2)}_{\mu \nu}|_{\varphi = 0} \cr} \right), 
\label{Tmatrix}
\ea
where 
\ba
T^{(1)}_{\mu \nu}|_{\varphi = 0} 
\A = \A - \ \epsilon_{\lambda \rho \sigma \nu} 
        (\overline{\tilde \psi}_R^{\lambda} 
        \gamma_{\mu} \partial^{\rho} \psi_R^{\sigma} 
        - \overline{\tilde \psi}_L^{\lambda} 
        \gamma_{\mu} \partial^{\rho} \psi_L^{\sigma}) \nonu
\A = \A \ \epsilon_{\lambda \rho \sigma \nu} 
        \overline{\tilde \psi}^{\lambda} \gamma_5 
        \gamma_{\mu} \partial^{\rho} \psi^{\sigma}, 
\label{T1-p0} \\
T^{(2)}_{\mu \nu}|_{\varphi = 0} 
\A = \A -i \ \epsilon_{\lambda \rho \sigma \nu} 
        (\overline{\tilde \psi}_R^{\lambda} 
        \gamma_{\mu} \partial^{\rho} \psi_R^{\sigma} 
        + \overline{\tilde \psi}_L^{\lambda} 
        \gamma_{\mu} \partial^{\rho} \psi_L^{\sigma}) \nonu
\A = \A -i \ \epsilon_{\lambda \rho \sigma \nu} 
        \overline{\tilde \psi}^{\lambda} 
        \gamma_{\mu} \partial^{\rho} \psi^{\sigma}. 
\label{T2-p0}
\ea
Thus, $T^{(1)}_{\mu \nu}$ and $T^{(2)}_{\mu \nu}$ 
are mixed as the parameter $\varphi$ changes. 
Since both of $\psi_{\mu}$ and $\tilde \psi_{\mu}$ 
are Majorana spinors, the $T^{(1)}_{\mu \nu}|_{\varphi = 0}$ 
of (\ref{T1-p0}) can be diagonalized as 
\be
T^{(1)}_{\mu \nu}|_{\varphi = 0}
= \epsilon_{\lambda \rho \sigma \nu} 
  (\overline \psi^{1 \lambda} \gamma_5 
  \gamma_{\mu} \partial^{\rho} \psi^{1 \sigma} 
  - \overline \psi^{2 \lambda} \gamma_5 
  \gamma_{\mu} \partial^{\rho} \psi^{2 \sigma}), 
\label{T1-diag}
\ee
where $\psi^1_{\mu} 
:= (1/2) (\psi_{\mu} + \tilde \psi_{\mu})$ and 
$\psi^2_{\mu} 
:= (1/2) (\psi_{\mu} - \tilde \psi_{\mu})$. 
The minus sign in (\ref{T1-diag}) 
means the appearance of negative energy: 
Namely, the positivity of $T^{(1)}_{\mu \nu}|_{\varphi = 0}$ 
is not guaranteed. 
On the other hand, the $T^{(2)}_{\mu \nu}|_{\varphi = 0}$ 
of Eq.(\ref{T2-p0}) is not diagonalized by using 
$\psi^1_{\mu}$ and $\psi^2_{\mu}$. 
However we have 
\ba
\A \A T^{(2)}_{\mu \nu}|_{\varphi = 0}[\psi, \tilde \psi] 
  = T^{(1)}_{\mu \nu}|_{\varphi = \pi/2}[\psi, \tilde \psi] 
  = T^{(1)}_{\mu \nu}|_{\varphi = 0}[\psi', \tilde \psi], \\
\A \A T^{(1)}_{\mu \nu}|_{\varphi = 0}[\psi, \tilde \psi] 
  = - T^{(2)}_{\mu \nu}|_{\varphi = \pi/2}[\psi, \tilde \psi] 
  = - T^{(2)}_{\mu \nu}|_{\varphi = 0}[\psi', \tilde \psi], 
\ea
where we define $\psi'_{\mu} := 
{\rm exp} \{ (i \pi \gamma_5)/2 \} \psi_{\mu}$. 
Therefore $T^{(2)}_{\mu \nu}|_{\varphi = 0}[\psi, \tilde \psi]$ 
is not positive definite, either. 

If we impose the reality condition, 
$b_{(\mu \nu)} = 0$ and $\overline{\tilde \psi}_{\mu} 
= \overline \psi_{\mu}$, together with $\varphi = 0$, 
then the $S^*_{\mu \nu \rho}$ of (\ref{S-star}) 
can be calculated as 
\be
S^*_{\mu \nu \rho} 
= -i {\epsilon^{\alpha \beta}} \! _{\rho [\mu} 
  \overline \psi_{|\alpha|} \gamma_{\nu]} \psi_{\beta}, 
\ee
by virtue of which the symmetrized energy-momentum tensor 
$T^{(2) ({\rm sym})}_{\mu \nu}$ of (\ref{T2-sym}) 
vanishes and the positivity of $T^{(1) ({\rm sym})}_{\mu \nu}$ 
is recovered.

\newpage
\newsection{Local SUSY invariant Lagrangian 
to order $\kappa^0$}

In the usual $N = 1$ SUGRA, the full nonlinear theory 
can be constructed from the linearized (free) theory, 
making the rigid SUSY transformations local 
and adding appropriate terms to the free Lagrangian 
order-by-order in the gravitational constant $\kappa$ 
\cite{West,FNF}. For example, one can construct 
a local SUSY invariant Lagrangian to order $\kappa^0$ 
by adding an interaction term 
proportional to the energy-momentum tensor. 
Similarly, we can obtain a local SUSY invariant Lagrangian 
to order $\kappa^0$ also in $N = 1$ chiral SUGRA 
as we shall explain below. 
Hereafter we shall write the $\kappa$ explicitly. 
In the linearized theory, the free Lagrangian of $N = 1$ 
chiral SUGRA is twice the usual $N = 1$ SUGRA: 
Namely, taking $\varphi = 0$ for simplicity, 
the free field limit of the real Lagrangian density, 
${\cal L} = {\cal L}_G + {\cal L}_{{\rm RS}}$, is 
\be
L^0 = L^0_G + L^0_{{\rm RS}}, 
\label{Lfree}
\ee
where the linearized gravitational Lagrangian, 
$L^0_G$, is 
\be
L^0_G = - (a^{(\mu \nu)} G_{\mu \nu}[a] 
          - b^{(\mu \nu)} G_{\mu \nu}[b]) 
\ee
and the free Lagrangian of (Majorana) Rarita-Schwinger 
fields, $L^0_{{\rm RS}}$, is 
\be
L^0_{{\rm RS}} = \epsilon^{\mu \nu \rho \sigma} 
        \overline {\tilde \psi}_{\mu} \gamma_5 \gamma_\rho 
        \partial_\sigma \psi_{\nu}, 
\ee
which can be diagonalized as 
\be
L^0_{{\rm RS}} = \epsilon^{\mu \nu \rho \sigma} 
              (\overline \psi_{\mu}^1 \gamma_5 \gamma_\rho 
               \partial_\sigma \psi_{\nu}^1 
             - \overline \psi_{\mu}^2 \gamma_5 \gamma_\rho 
               \partial_\sigma \psi_{\nu}^2) 
\ee
up to a total divergence term, 
where $\psi^1_{\mu}$ and $\psi^2_{\mu}$ are defined 
below (\ref{T1-diag}). 

The linearized theory of $N = 1$ chiral SUGRA 
possesses local gauge invariance and rigid SUSY invariance 
just like the usual $N = 1$ SUGRA: Indeed, 
the $L^0$ of (\ref{Lfree}) is invariant 
under local gauge transformations, 
\ba
\A \A \delta a_{(\mu \nu)} 
      = \partial_{\mu} \xi_{\nu}(x) 
      + \partial_{\nu} \xi_{\mu}(x), \nonu
\A \A \delta b_{(\mu \nu)} 
      = \partial_{\mu} \eta_{\nu}(x) 
      + \partial_{\nu} \eta_{\mu}(x), \nonu
\A \A \delta \psi_{\mu} 
      = \partial_{\mu} \epsilon(x), \nonu
\A \A \delta \tilde \psi_{\mu} 
      = \partial_{\mu} \tilde \epsilon(x). 
\label{gauge}
\ea
In contrast with the usual $N = 1$ SUGRA, 
on the other hand, $L^0$ is invariant 
under two kinds of rigid SUSY transformations. 
One is the following rigid SUSY transformations 
with supersymmetric partners being $(a_{(\mu \nu)}, \psi^1_{\mu})$ 
and $(b_{(\mu \nu)}, \psi^2_{\mu})$: 
\ba
\A \A \delta a_{(\mu \nu)} 
      = -i \overline \psi^1_{(\mu} \gamma_{\nu)} \alpha^1, \ \ 
      \delta \psi^1_{\mu} 
      = -2i S^{\rho \sigma} \partial_{\rho} a_{(\sigma \mu)} 
      \alpha^1, \nonu
\A \A \delta b_{(\mu \nu)} 
      = -i \overline \psi^2_{(\mu} \gamma_{\nu)} \alpha^2, \ \ 
      \delta \psi^2_{\mu} 
      = -2i S^{\rho \sigma} \partial_{\rho} b_{(\sigma \mu)} 
      \alpha^2, 
\label{tw-SUSY}
\ea
where $\alpha^1$ and $\alpha^2$ are 
constant Majorana spinor parameters. 
In this case, if we make $\alpha^1$ and $\alpha^2$ local 
and add appropriate terms to the $L^0$ of (\ref{Lfree}) 
order-by-order in $\kappa$, 
then the resultant Lagrangian density is twice 
that of the usual $N = 1$ SUGRA: 
Namely, we have ${\cal L} = {\cal L}^1_{N = 1 {\rm SUGRA}} 
- {\cal L}^2_{N = 1 {\rm SUGRA}}$. 

The other is the rigid SUSY transformations, 
\ba
\A \A \delta a_{(\mu \nu)} 
      = -{i \over 2} 
      (\overline{\tilde \psi}_{(\mu} \gamma_{\nu)} \alpha 
      + \overline \psi_{(\mu} \gamma_{\nu)} \tilde \alpha), 
      \nonu
\A \A \delta b_{(\mu \nu)} 
      = {1 \over 2} 
      (\overline{\tilde \psi}_{(\mu} \gamma_5 
      \gamma_{\nu)} \alpha 
      - \overline \psi_{(\mu} \gamma_5 
      \gamma_{\nu)} \tilde \alpha), \nonu
\A \A \delta \psi_{\mu} 
      = -2i S^{\rho \sigma} (\partial_{\rho} a_{(\sigma \mu)} 
      + i \gamma_5 \ \partial_{\rho} b_{(\sigma \mu)}) \alpha, 
      \nonu
\A \A \delta \tilde \psi_{\mu} 
      = -2i S^{\rho \sigma} (\partial_{\rho} a_{(\sigma \mu)} 
      - i \gamma_5 \ \partial_{\rho} b_{(\sigma \mu)}) 
      \tilde \alpha 
\label{r-SUSY}
\ea
with $\alpha$ and $\tilde \alpha$ being 
constant Majorana spinor parameters. 
Although the gauge and SUSY transformations 
(\ref{gauge}) and (\ref{r-SUSY}) 
form a closed algebra on shell, 
the rigid SUSY transformations (\ref{r-SUSY}) 
are not twice the usual $N = 1$ SUGRA. 

Let us construct a local SUSY invariant Lagrangian 
to order $\kappa^0$. If the spinor parameters 
$\alpha$ and $\tilde \alpha$ in (\ref{r-SUSY}) 
become spacetime dependent, i.e., $\alpha = \alpha(x)$ 
and $\tilde \alpha = \tilde \alpha(x)$, 
then the $L^0$ of (\ref{Lfree}) is no longer invariant: 
The variation of $L^0$ can be expressed as 
\be
\delta L^0 
= (\partial_{\mu} \overline{\tilde \alpha}) 
(J^{1 \mu}[a, \psi] + J^{2 \mu}[b, \psi]) 
+ (\partial_{\mu} \overline \alpha) 
(\tilde J^{1 \mu}[a, \tilde \psi] 
+ \tilde J^{2 \mu}[b, \tilde \psi]), 
\label{del-L0}
\ee
up to total divergence, 
because $L^0$ is invariant when $\alpha$ and $\tilde \alpha$ 
are constant. However, $J^{1 \mu}[a, \psi]$, 
$J^{2 \mu}[b, \psi]$, $\tilde J^{1 \mu}[a, \tilde \psi]$ 
and $\tilde J^{2 \mu}[b, \tilde \psi]$ are not uniquely fixed: 
As example, we shall explain how the ambiguity 
arises in $J^{1 \mu}[a, \psi]$. 
The variation of $L^0$ with respect to the transformations 
(\ref{r-SUSY}) is 
\be
\delta L^0 
= (\partial_{\mu} a_{(\lambda \nu)}) 
\epsilon^{\mu \nu \rho \sigma} \overline{\tilde \alpha} 
\gamma_5 \gamma^{\lambda} \partial_{\rho} \psi_{\sigma} 
+ \cdots. 
\label{del-L0J1}
\ee
If the spinor parameter $\tilde \alpha$ is constant, 
the first term of (\ref{del-L0J1}) is just a total divergence. 
However, when $\tilde \alpha$ is spacetime dependent, 
this term can be rewritten as 
\ba
\A \A (\partial_{\mu} a_{(\lambda \nu)}) 
      \epsilon^{\mu \nu \rho \sigma} \overline{\tilde \alpha} 
      \gamma_5 \gamma^{\lambda} \partial_{\rho} \psi_{\sigma} \nonu
\A \A = (\partial_{\mu} \overline{\tilde \alpha}) 
      \{ - p a_{(\lambda \nu)} \epsilon^{\mu \nu \rho \sigma} 
      \gamma_5 \gamma^{\lambda} \partial_{\rho} \psi_{\sigma} 
      + (1 - p) (\partial_{\rho} a_{(\lambda \nu)}) 
      \epsilon^{\mu \nu \rho \sigma} 
      \gamma_5 \gamma^{\lambda} \psi_{\sigma} \} \nonu
\A \A \ \ \ + \ [{\rm a \ total \ divergence}] 
\label{first}
\ea
with $p$ being an arbitrary real constant. 
Although the two terms proportional to $p$ 
in (\ref{first}) are combined to a total divergence, 
they do lead to non-trivial ambiguity in $J^{1 \mu}[a, \psi]$. 
We can explicitly show that the remaining terms 
of (\ref{del-L0J1}) do not lead to any ambiguity 
in $J^{1 \mu}[a, \psi]$. Thus we have 
\be
J^{1 \mu}[a, \psi] 
      = - p a_{(\lambda \nu)} \epsilon^{\mu \nu \rho \sigma} 
      \gamma_5 \gamma^{\lambda} \partial_{\rho} \psi_{\sigma} 
      + (1 - p) (\partial_{\rho} a_{(\lambda \nu)}) 
      \epsilon^{\mu \nu \rho \sigma} 
      \gamma_5 \gamma^{\lambda} \psi_{\sigma} + \cdots. 
\label{J1} 
\ee
Note that similar ambiguity also appears in the usual $N = 1$ SUGRA. 
In the same manner, we get 
\ba
\A \A J^{2 \mu}[b, \psi] 
      = iq b_{(\lambda \nu)} \epsilon^{\mu \nu \rho \sigma} 
      \gamma^{\lambda} \partial_{\rho} \psi_{\sigma} 
      - i(1 - q) (\partial_{\rho} b_{(\lambda \nu)}) 
      \epsilon^{\mu \nu \rho \sigma} 
      \gamma^{\lambda} \psi_{\sigma} + \cdots, \\
\A \A \tilde J^{1 \mu}[a, \tilde \psi] 
      = - p' a_{(\lambda \nu)} \epsilon^{\mu \nu \rho \sigma} 
      \gamma_5 \gamma^{\lambda} \partial_{\rho} 
      \tilde \psi_{\sigma} 
      + (1 - p') (\partial_{\rho} a_{(\lambda \nu)}) 
      \epsilon^{\mu \nu \rho \sigma} 
      \gamma_5 \gamma^{\lambda} 
      \tilde \psi_{\sigma} + \cdots, \\
\A \A \tilde J^{2 \mu}[b, \tilde \psi] 
      = - iq' b_{(\lambda \nu)} \epsilon^{\mu \nu \rho \sigma} 
      \gamma^{\lambda} \partial_{\rho} \tilde \psi_{\sigma} 
      + i(1 - q') (\partial_{\rho} b_{(\lambda \nu)}) 
      \epsilon^{\mu \nu \rho \sigma} 
      \gamma^{\lambda} \tilde \psi_{\sigma} + \cdots 
      \label{tJ2}
\ea
with $p'$, $q$ and $q'$ being arbitrary 
real constants. 

Inspection of (\ref{del-L0}) shows that the invariance 
under transformations with local spinor parameter 
$\tilde \alpha(x)$ is recovered to order $\kappa^0$ 
if the interaction term, $(-\kappa/2) \overline{\tilde \psi}_{\mu} 
(J^{1 \mu}$ $[a, \psi] + J^{2 \mu}[b, \psi])$, 
is added to $L^0$, and if we simultaneously make 
the gauge transformation of $\tilde \psi_{\mu}$ 
with $\tilde \epsilon(x) = (2/\kappa) \tilde \alpha(x)$. 
On the other hand, the interaction term, 
$(-\kappa/2) \overline \psi_{\mu} (\tilde J^{1 \mu}[a, \tilde \psi] 
+ \tilde J^{2 \mu}[b, \tilde \psi])$, is needed 
if the spinor parameter $\alpha$ becomes spacetime dependent. 
Since these two interaction terms must be the same, 
we require 
\be
\overline{\tilde \psi}_{\mu} 
(J^{1 \mu}[a, \psi] + J^{2 \mu}[b, \psi]) 
= \overline \psi_{\mu} 
(\tilde J^{1 \mu}[a, \tilde \psi] 
+ \tilde J^{2 \mu}[b, \tilde \psi]). 
\label{J=tJ}
\ee
Then the constants $p$, $p'$, $q$ and $q'$ 
are uniquely determined as $p = p' = 2$ and $q = q' = 2$. 
Since $J^{1 \mu}[a, \psi]$, $J^{2 \mu}[b, \psi]$, 
$\tilde J^{1 \mu}[a, \tilde \psi]$ and 
$\tilde J^{2 \mu}[b, \tilde \psi]$ are now fixed 
and satisfies (\ref{J=tJ}), we modify the Lagrangian as 
\be
L^1 = L^0 
- {\kappa \over 2} \overline{\tilde \psi}_{\mu} 
(J^{1 \mu}[a, \psi] + J^{2 \mu}[b, \psi]), 
\label{Lint}
\ee
which is invariant to order $\kappa^0$ 
under the local SUSY transformations combined 
with the gauge transformations with the identification, 
$\epsilon(x) = (2/\kappa) \alpha(x)$ 
and $\tilde \epsilon(x) = (2/\kappa) \tilde \alpha(x)$. 

The interaction term in the $L^1$ of (\ref{Lint}) 
can be rewritten as 
$\kappa (a^{(\mu \nu)} T^{(1) ({\rm sym})}_{\mu \nu}
+ b^{(\mu \nu)} T^{(2) ({\rm sym})}_{\mu \nu})$. 
Thus (\ref{leq-Ta}) and (\ref{leq-Tb}) are derived 
from $L^1$ by taking variation with respect to $a^{(\mu \nu)}$ 
and $b^{(\mu \nu)}$ respectively. 
It should also be noted that the correction term of $L^1$ 
is not twice the usual $N = 1$ SUGRA, 
because $\psi_{\mu}$ and $\tilde \psi_{\mu}$ are contained 
both in $T^{(1)}_{\mu \nu}$ and $T^{(2)}_{\mu \nu}$ 
as shown in (\ref{T1-p0}) and (\ref{T2-p0}).

\newsection{Conclusion}

We have studied the structure of linearized field 
equations in the chiral formulation of gravity 
with the complex tetrad, 
and seen how the real and imaginary parts 
of the complex tetrad are coupled to matter fields 
in the weak field approximation. 
As an example, the explicit form of energy-momentum tensors 
has been calculated for (Majorana) Rarita-Schwinger fields. 
Starting from the linearized (free) theory 
of $N = 1$ chiral SUGRA, we have then obtained the Lagrangian 
which is invariant under local SUSY transformations 
to order $\kappa^0$. 
The resultant Lagrangian just reproduces the field equations 
in the weak field approximation. 
We expect that the full nonlinear theory 
of $N = 1$ chiral SUGRA can be constructed 
by adding appropriate terms order-by-order in $\kappa$ 
to the first-order Lagrangian we have obtained. 
We are also trying to introduce additional auxiliary fields 
into the linearized theory of $N = 1$ chiral SUGRA 
as the preliminary step to construct the full nonlinear 
theory with auxiliary fields of $N = 1$ chiral SUGRA.

\bigskip
\bigskip
\bigskip

\begin{flushleft}
{\large{\bf Acknowledgements}}
\end{flushleft}

\bigskip

We would like to thank the members of Physics Department 
at Saitama University 
for discussions and encouragement, 
Professor Y. Tanii and other members.

\appendix{Appendix}{1}

In this appendix, we derive identities 
from general coordinate and local Lorentz invariances 
of the chiral Lagrangian density 
\be
{\cal L}^{(+)} = {\cal L}^{(+)} [q_{\mu}; q_{\mu,\nu}; 
             e_{k \mu}; e_{k \mu,\nu}; 
             A^{(+)}_{ij \mu}; A^{(+)}_{ij \mu,\nu}], 
\ee
where matter fields are collectively denoted by $q_{\mu}$ 
\cite{HB,NSH}. 
An arbitrary variation of ${\cal L}^{(+)}$ is 
\be
\delta {\cal L}^{(+)} 
= {{\delta {\cal L}^{(+)}} \over {\delta F}} \overline \delta F 
  + \partial_{\nu} \left( {{\partial {\cal L}^{(+)}} \over 
  {\partial F_{,\nu}}} \overline \delta F 
  + {\cal L}^{(+)} \delta x^{\nu} \right), 
\ee
where $F$ denotes $q_{\mu}, e_{k \mu}$ and $A^{(+)}_{ij \mu}$ 
collectively, and $\overline \delta$ is the Lie derivative 
defined by 
$\overline \delta F = \delta F - F_{,\nu} \delta x^{\nu}$. 

If the ${\cal L}^{(+)}$ is invariant 
under general coordinate and local Lorentz transformations, 
we have Noether's identity, 
\be
{{\delta {\cal L}^{(+)}} \over {\delta F}} \overline \delta F 
+ \partial_{\nu} \left( {{\partial {\cal L}^{(+)}} \over 
{\partial F_{,\nu}}} \overline \delta F 
+ {\cal L}^{(+)} \delta x^{\nu} \right) \equiv 0. 
\label{Nid}
\ee

For complex local Lorentz transformations, 
variation of the fields is given by 
\ba
\A \A \overline \delta q_{\mu} 
      = {i \over 2} \epsilon_{ij} S^{ij} q_{\mu}, \ \ 
      \overline \delta e_{k \mu} 
      = {\epsilon_k}^l e_{l \mu} \nonu
\A \A \overline \delta A^{(+)}_{ij \mu} 
      = {\epsilon_i}^k A^{(+)}_{kj \mu} 
      + {\epsilon_j}^k A^{(+)}_{ik \mu} 
      - \epsilon^{(+)}_{ij,\mu} 
\label{vari-L}
\ea
where $\epsilon_{ij} = \epsilon_{[ij]}$ 
is an arbitrary complex parameter. 
Here $\epsilon^{(+)}_{ij} := (1/2) \{ \epsilon_{ij} 
- (i/2) {\epsilon_{ij}} \! ^{kl} \epsilon_{kl} \}$, 
and it is independent 
of $\epsilon^{(-)}_{ij} := (1/2) \{ \epsilon_{ij} 
+ (i/2) {\epsilon_{ij}} \! ^{kl} \epsilon_{kl} \}$. 
Using (\ref{vari-L}) in (\ref{Nid}) 
with ${\cal L}^{(+)} = {\cal L}^{(+)}_M$ yields 
\be
{1 \over 2} \left( {\bf T}^{(+) [ij]} - {i \over 2} 
{\epsilon^{ij}} \! _{kl} {\bf T}^{(+) kl} \right) 
- {1 \over 2} D^{(+)}_{\nu} {\bf S}^{(+)ij \nu} 
+ {{\delta {\cal L}^{(+)}_M} \over {\delta q_{\mu}}} 
{i \over 2} S^{(+) ij} q_{\mu} \equiv 0 
\label{Nid-L1}
\ee
and 
\be
{1 \over 2} \left( {\bf T}^{(+) [ij]} + {i \over 2} 
{\epsilon^{ij}} \! _{kl} {\bf T}^{(+) kl} \right) 
+ {{\delta {\cal L}^{(+)}_M} \over {\delta q_{\mu}}} 
{i \over 2} S^{(-) ij} q_{\mu} \equiv 0 
\label{Nid-L2}
\ee
with ${\bf T}^{(+)}_{ij} = e_{j \mu} {{{\bf T}^{(+)}}_i}^{\mu}$. 
Here ${{{\bf T}^{(+)}}_i}^{\mu}$ and ${\bf S}^{(+)ij \mu}$ 
denotes  $e \ {{T^{(+)}}_i}^{\mu}$ and $e \ S^{(+)ij \mu}$ 
with ${{T^{(+)}}_i}^{\mu}$ and $S^{(+)ij \mu}$ 
being defined by (\ref{EM+}) and (\ref{spin+}) respectively. 
Similarly, the identity for general coordinate invariance 
of ${\cal L}^{(+)}_M$ can be written as 
\be
\partial_{\nu} {{{\bf T}^{(+)}}_{\mu}}^{\nu} 
- {e^i}_{\nu,\mu} {{{\bf T}^{(+)}}_i}^{\nu} 
- A^{(+)}_{ij \mu} {\bf T}^{(+) ij} 
+ {1 \over 2} R^{(+)}_{ij \mu \nu} {\bf S}^{(+)ij \nu} 
- {{\delta {\cal L}^{(+)}_M} \over {\delta q_{\nu}}} 
D^{(+)}_{\mu} q_{\nu} 
+ \partial_{\nu} 
({{\delta {\cal L}^{(+)}_M} \over {\delta q_{\nu}}} q_{\mu}) 
\equiv 0, 
\label{Nid-G}
\ee
by using (\ref{Nid-L1}). 

With the help of matter field equations, 
the identities (\ref{Nid-L1}) and (\ref{Nid-L2}) 
become, in the special relativistic limit, 
\be
T^{(+)}_{[\mu \nu]} - {i \over 2} 
      {\epsilon_{\mu \nu}} \! ^{\rho \sigma} T^{(+)}_{\rho \sigma} 
      \equiv \partial^{\rho} S^{(+)}_{\mu \nu \rho}, \ \ 
T^{(+)}_{[\mu \nu]} + {i \over 2} 
      {\epsilon_{\mu \nu}} \! ^{\rho \sigma} T^{(+)}_{\rho \sigma} 
      \equiv 0, 
\label{Tet+}
\ee
while (\ref{Nid-G}) gives the conservation law, 
\be
\partial^{\nu} T^{(+)}_{\mu \nu} \equiv 0. 
\label{conserv+}
\ee
Since ${{T^{(+)}}_i}^{\mu}$ and $S^{(+)ij \mu}$ 
are related to ${{T^{(1)}}_i}^{\mu}$, ${{T^{(2)}}_i}^{\mu}$ 
and $S^{ij \mu}$ by (\ref{rel-T}) and (\ref{rel-S}), 
we obtain the conservation law (\ref{conserv}) 
from (\ref{conserv+}), 
and further the relations (\ref{Tet}) and (\ref{T1-T2}) 
from (\ref{Tet+}).



\end{document}